\documentclass[12pt]{article}
\begin{document}
\begin{center}
{\bf Comment on evidence for new interference phenomena in the decay
$D^+\to K^- \pi^+ \mu^+ \nu$}
\vskip 1cm
B. Ananthanarayan\\
Centre for High Energy Physics\\
Indian Institute of Science\\
Bangalore 560 012, India\\
\bigskip
K. Shivaraj\\
Department of Physics,\\
Indian Institute of Science\\
Bangalore 560 012, India\\
\end{center}

\bigskip

\begin{abstract}
The experimental determination of low energy $\pi K$ scattering phase shifts
would assist in determining scattering lengths as well as low energy
constants of chiral perturbation theory for which sum rules have been
constructed.  The FOCUS collaboration has presented evidence for
interference pheomena from their analysis of $D_{l4}$ decays based on
decay amplitudes suitable for a cascade decay $D\to K^*\to K \pi$.  We
point out that if the well-known full five body kinematics are taken 
into account, $\pi K$ scattering phases may be extracted.
We also point out that other distributions considered
in the context of $K_{l4}$ decays
can be applied to charm meson decays to provide constraints on
violation of $|\Delta I|=1/2$ rule and T-violation.
\end{abstract}

\begin{small}
\end{small}

\newpage

\noindent {\bf 1.} Chiral perturbation theory~\cite{GL} 
as the low energy effective theory
of the standard model is now in a remarkably mature phase.  Several
processes have been computed to two-loop accuracy and remarkable predictions
exist for low energy processes.  One of the important processes that
has been studied is that of $\pi\pi$ scattering,
for a recent comprehensive review, see ref.~\cite{CGL}.  
It has been traditionally
difficult to study this experimentally in the low-energy regime due to
the absence of pion targets.  One important source of information comes from
the rare kaon decay $K_{l4}$.  Using well-known techniques~\cite{CM,PT} 
one can extract the phase difference for pion scattering of the iso-scalar
S-wave and iso-triplet P-wave phase shifts $\delta^0_0-\delta^1_1$ from
an analysis of the angular distributions, where the final state or
Watson theorem relates the phase of the decay form factors to the scattering
phase shifts.  Recently the E865 collaboration~\cite{E865} at 
Brookhaven National Laboratory has carried out the analysis of data from a 
high statistics experiments which has brought about a remarkable marriage 
between experiment and theory.  
There are preliminary measurements also from NA48 for the semi-leptonic
decays, ref.~\cite{NA48}.
Scattering lengths will also be measured
at high precision by the CERN experiment DIRAC from the lifetime of
the pionium atom, and from the enormous statistics gathered by the
NA48 collaboration by employing the recent proposal of Cabibbo, see
refs.~\cite{Cabibbo,CI}, of analyzing the cusp structure of the
invariant mass of the dipion system produced in the reaction
$K^+\to \pi^+ \pi^+ \pi^-$.

\bigskip

\noindent {\bf 2.} 
Chiral perturbation theory that involves the s-quark degree of
freedom is yet to be tested at a corresponding degree of precision.
One sensitive laboratory is the pion-Kaon scattering amplitude~\cite{BKM}.  
For recent studies on the comparison between the amplitudes 
evaluated in chiral perturbation theory, and phenomenologcal
determination, see refs.~\cite{AB,ABM}.  It has
been pointed in these that it is desirable to have
high precision phase shift determinations 
so that accurate predictions for scattering lengths can be made.
The search for an experimental system where these
phase shifts can be measured, leads us naturally to an analog of the 
$K_{l4}$ decay in the
charm-meson system, which is the decay $D_{l4}$\footnote{Indirect sources
of information include pion production from scattering of kaons off
nuclei, e.g.~\cite{Aston}.}.
It is clear that one might be
able to extract information on the $\pi K$ scattering amplitude as well
due to the final state or Watson theorem.  What is required is an analog
for the technique used in the case of
$K_{l4}$ decay for the $D_{l4}$ decay.  Note that in the $D_{l4}$ case,
the dimeson pair in the final state is composed of unequal mass particles 
and that the
iso-spin of the system is different from that in the corresponding $\pi\pi$
system.  At leading order in the weak interaction, one obtains only 
$|\Delta I|=1/2$ amplitudes and the system yields information on the phase
shift difference $\delta^{1/2}_0-\delta^{1/2}_1$.  A comprehensive
and self-contained account of this is to be found in ref.~\cite{KKPS}.
Note that for the moment, the analog of the pionium system for
the $\pi K$ atom is only in the planning stage, and 
determination of the $\pi K$ scattering lengths from an analog
of the proposal of Cabibbo from, say 
$D^+ \to K^- \pi^+ \pi^+$  or  $\bar{K}^0 \pi^+ \pi^0$
would not be feasible due to limited statistics. 
As a result, it is imperative that the $D_{l4}$ decay be
exploited to determine the phase shifts of interest.
The data so obtained could be in conjuction with the 
recent accurate solutions to the Roy-Steiner equations~\cite{BDM}.

\bigskip

\noindent{\bf 3.}  In this paragraph, we briefly recall the main features
of the formalism of ref.~\cite{KKPS}.  
The process considered is
\begin{equation}
D(p_1)\to K(p_2) +\pi(p_3)+l(k)+\nu(k').
\end{equation}
The authors give an explicit form for the 5-fold differential width
\begin{equation}\label{5fold}
{d^5\Gamma \over dq^2 ds_{23}d\cos\theta d\chi d\cos\theta^*}
={G_F^2 |V_{cs}|^2 q^2 \sqrt{a_2} X \over 96 (2 \pi)^6 m_1^3} \sum_i l_i H_i,
\end{equation} 
where $q=k+k'$, $m_1$ is the mass of the $D$ meson, $s_{23}=(p_2+p_3)^2$,
$a_2=4 |{\bf p_2}|^2/s_{23}$, $X=\sqrt{s_{23}}|{\bf p_1}|$,
$\theta$ is the angle between the charged lepton and the $D$ meson in
the dilepton center of mass frame, 
$\theta^*$ is the angle between the $K$ meson and
the $D$ meson in the dimeson center of mass frame, $\chi$ is the angle between
the lepton and meson decay planes, and the sum over $i$ runs over
the symbols $U,\, L,\, T,\, V,\, P,\, F,\, I,\, N,\, A$, with the
$H_i$ being the helicity structure functions, and the
$l_i$ given as follows for the case of massless charged leptons:
\begin{eqnarray*}
& \displaystyle
l_U={3\over 8} (1+\cos^2\theta), \, l_L={3\over 4} \sin^2\theta, \,
l_T={3\over 4} \sin^2\theta \cos(2\chi), & \\
& \displaystyle
l_V=-{3\over 4} \sin^2\theta \sin(2\chi),\, l_P={3\over 4} \cos\theta, \,
l_F={3\over 2\sqrt{2}} \sin(2\theta) \sin\chi, & \\
& \displaystyle
l_I=-{3\over 2 \sqrt{2}} \sin(2\theta) \cos\chi, \, l_N={3\over \sqrt{2}}
\sin \theta \sin \chi, \, l_A=-{3\over \sqrt{2}} \sin\theta \cos\chi. &
\end{eqnarray*}
We do not explicitly list all the $H_i$ except for a few for
purposes of illustration (see below).

Writing the hadronic matrix element as
\begin{eqnarray*}
& \displaystyle
\langle p_2,p_3 | A_\mu + V_\mu | p_1 \rangle= & \\
& \displaystyle
{1\over m_1}\left[f(p_2+p_3)_\mu +g(p_2-p_3)_\mu + r q_\mu+
{i h\over m_1^2}\epsilon_{\mu\nu\alpha\beta} q^\nu (p_2+p_3)^\alpha 
(p_2-p_3)^\beta\right],
\end{eqnarray*}
where the form factors
$f,\, g,\, r,\, {\rm and}\, h$ are in general functions of 
$s_{23}$ ,$q^2$ and $\theta^*$ ($r$ makes no contribution in the case of
massless charged leptons).
The $H_i$ can now be expressed in terms of the form factors.
For instance,
\begin{eqnarray*}           
& \displaystyle
H_{F(A)}={X \over m_1^2}{\sqrt{a_2 s_{23}} \over \sqrt{2 q^2}m_1^2} 
{\rm Im\, (Re)} 
\left( h^*\left[Xf+gX{{m_2^2 - m_3^2}\over s_{23}}\right.\right. & \\
& \displaystyle
 \left.\left. +g\sqrt{a_2} {m_1^2-s_{23}-q^2\over 2}\cos\theta^*\right] 
\right)\sin\theta^*, & \\
& \displaystyle
H_V=-{X a_2 s_{23}\over m_1^4} {\rm Im}(h^* g) \sin^2\theta^*. &  
\end{eqnarray*}

It was shown first by Pais and Treiman that the choice of variables
made by Cabibbo and Maksymowicz leads to the simple decomposition,
eq.~(\ref{5fold}) of
the 5-fold differential width and thus makes the determination of
physical observables amenable.  Furthermore,
by parametrizing the functions $f,\, g,\, h$ and
identifying their phases with $\pi K$ phase shifts (a consequence of
Watson's theorem), the partial wave expansion of $f,\, g,\, {\rm and}\, h$ 
read
\begin{eqnarray*}
& \displaystyle
f = \tilde{f_s} e^{i\delta_0^{1/2}} + 
\tilde{f_p} e^{i\delta_1^{1/2}} \cos\theta^*+ \dots, & \\
& \displaystyle
g = \tilde{g_p} e^{i\delta_1^{1/2}} + \dots, & \\
& \displaystyle
h = \tilde{h_p} e^{i\delta_1^{1/2}} + \dots. & 
\end{eqnarray*}
It may, therefore be seen from the above that an analysis of
the decay distribution would yield information on the phase shifs
of interest.

\bigskip

\noindent {\bf 4.}  
The FOCUS Collaboration has recently published
``evidence for new interference phenomena in the decay
$D^+\to K^- \pi^+ \mu^+ \nu$''~\cite{Focus1}.  By including an
S-wave in a straightforward manner into the decay amplitude
that is dominated by the P-wave $K^*$ resonance, and finding
a superior fit to certain distributions, this result has been
established.  The decay amplitude has been adopted from ref.~\cite{KS}
which considers the three body final state kinematics.  
In particular, the process considered in~\cite{KS} is
the reaction of the type
\begin{equation}
D(p_1)\to K^*(p^*) +l(k) +\nu(k')
\end{equation}
for which the hadronic part of the amplitude is written down
in terms of the matrix element
\begin{equation}
\langle K^*(p^*)|A_\mu+V_\mu|D(p_1)\rangle =\epsilon_2^{*\alpha} T_{\mu\alpha},
\end{equation}
where
\begin{equation}
T_{\mu\alpha}= F_1^A g_{\mu\alpha}+F_2^A p_{1\mu} p_{1\alpha} + F_3^A
q_\mu p_{1\alpha} + i F^V \epsilon_{\mu\alpha\rho\sigma} p_1^\rho p^{*\sigma},
\end{equation}
and $q_\mu=(p_1-p^*)_\mu$ is the momentum transfer.  
Note that $F_3^A$ contributes
only in the case of decays with massive charged leptons.
The differential decay rates are expressed in terms of 
helicity amplitudes which evaluate to
\begin{eqnarray}
& \displaystyle
H_0={1\over 2 M_1 \sqrt{q^2}}
\left( (M_1^2-M^{*2}-q^2) F_1^A + 2 M_1^2 p^2 F_2^A
\right), & \nonumber \\
& \displaystyle
H_\pm=F_1^A\pm M_1 p F^V, & \nonumber
\end{eqnarray}
where $p$ is the momentum of the $K^*$ in the $D$ rest system,
$M_1$ and $M^*$ are the masses of the $D$ and the $K^*$ respectively.
In ref.~\cite{Focus1}\footnote{
In ref.~\cite{KS} a discussion is provided on the 
multipole behaviour that is expected of the
functions $F^A_1,\, F^A_2,\ F_V$; the
FOCUS collaboration assumes all of them to have
monopole behaviour in their analysis, ref.~\cite{Focus1}.}, 
the expressions are provided for the
massless lepton case,
for which case the differential decay rate is written down as:
\begin{eqnarray}
& \displaystyle
{d^4\Gamma(D\to K^*\to K\pi)\over dq^2 d\cos\theta d\chi d\cos\theta^*}
\propto  B(K^*\to K \pi)& \nonumber \\
& \displaystyle
{9\over32}\left((1+\cos^2\theta)\sin^2\theta^*(|H_+|^2+|H_-|^2)+
      4 \sin^2\theta\cos^2\theta^*|H_0|^2-  \right. & \nonumber \\
& \displaystyle \left. 
      2\sin^2\theta \cos 2\chi \sin^2\theta^*{\rm Re}(H_+ H_-^*) -
 \sin 2\theta \cos\chi \sin 2\theta^* {\rm Re}(H_+ H_0^*+H_- H_0^*)
+\right. & \nonumber \\
& \displaystyle \left.
2\cos\theta \sin^2 \theta^*(|H_+|^2-|H_-|^2)-
2 \sin\theta \cos\chi \sin 2 \theta^* {\rm Re}(H_+ H_0^*-H_- H_0^*)
\right). & \nonumber
\end{eqnarray}
We note here that in the above,
(a) the result is expressed in the notation of ref.~\cite{KS}
along with the assumptions stated therein on contributions proportional
to ${\rm Im} H_i H^*_j,\, i\neq j$, (b)
and taking into account the remarks given in the last paragraph of
Sec. 4 of ref.~\cite{KKPS}, and (c)
and also that the FOCUS collaboration has taken lepton mass effects
into account for the results presented in ref.~\cite{Focus1}.

Note that in the treatment above, there will be no contributions of
the type $i=F,\, N,\, V$.   The FOCUS collaboration in the analysis of
its data, finds that a simple analysis based on a $1^{--}$ does
not fit the data well.  They make an {\it ad hoc} assumption and
introduce an amplitude with the properties of an S- wave $A \exp{i\delta}$.
Introducing this generates interference terms which would correspond
to terms that appear as $i=F,\, N$ and a term of the $i=L$ type.
This assumption cannot generate a term of the type 
$i=V$\footnote{Note that this is consistent with $H_V$ of the previous
paragraph vanishing in the S- and P- wave approximation}.
In~\cite{Focus1} the narrow-width approximation for the $K^*$ is
replaced by a Breit-Wigner and a full 5-fold distribution is written down.

\bigskip

\noindent {\bf 5.}  
It is our main comment here that the FOCUS collaboration must account
for the dynamics in its entireity by using the formalism of ref.~\cite{KKPS}.
In this manner, they would also be able to determine the phase shift
difference which would allow us to pin down low energy strong interactions
observables to better precision.  Note also that                                a complete description of the four body final state with
lepton mass effects included is presented
in ref.~\cite{KKPS}\footnote{In this regard, the FOCUS collaboration
has analyzed data with charged lepton mass effects with their 
modified formalism of ref.~\cite{KS} all along, and
present the relevant expressions in ref.~\cite{Focus2}.}.  
By binning the data
in the variable $s_{23}$ and carrying out integrations in the variables
$\theta^*$ and $\chi$ and fitting the resulting distribution
to experimental data, it would be possible to determine the
phase shifts and the form factors themselves. We note here that unlike in the 
$K_{l4}$ decay where the dimeson system is composed of equal mass particles, 
in the present case a ratio of e.g., $\langle H_F\rangle/
\langle H_A\rangle$ cannot directly 
yield information on $\delta_0^{1/2} -\delta_1^{1/2}$. Only a comprehensive fit 
to all the $\langle H_i\rangle$ can be used to extract this quantity.
In this regard, it would be useful to follow the procedure described
at length for the case of $K_{l4}$ decays in ref.~\cite{Rosselet}.

\bigskip

\noindent {\bf 6.}
We recall here that in the context of $K_{l4}$ decays, the original
5 body decay kinematics were discussed in ref.~\cite{CM}, where the authors
discussed only 1-dimensional distributions. 
In ref.~\cite{PT} 2-dimensional distributions were considered, and 
also analyzed in the context of limited statistics.  [The latter was the basis
of the analysis of the events from the well-known experiment,
ref.~\cite{Rosselet}.]  Subsequently Berends, Donnachie and
Oades (BDO)~\cite{BDO}, again considered 1-dimensional distributions, but
with limited statistics.  They also discussed $|\Delta I|=3/2, 5/2$
transitions, and also looked at tests of T-invariance. 
Recently the NA48 collaboration~\cite{NA48} has observed some evidence
for the violation of the $|\Delta I|=1/2$ rule consistent with
standard model expectations in $K_{l4}$ decays using the the
technique of BDO.

BDO in the context of $K_{l4}$ decays consider the 2-fold
distribution given by
\begin{eqnarray*}               
& \displaystyle
{d^2 \Gamma\over d\cos\theta^* d\chi} &
\end{eqnarray*}
which could receive contributions from T-violating interactions
assuming that higher wave contributions are absent.  Here we point out that
such a distribution for D-meson decays could receive
additional contributions from T-violation in the decays. 
Also considered in BDO are
the distributions
\begin{eqnarray*}
& \displaystyle {d\Gamma\over d\chi},\,  
{d\Gamma\over d\cos\theta^*} &
\end{eqnarray*}
which could be used to fit the form factors in an analysis independent
of the Pais-Treiman type distributions.
The work of BDO can be readily extended to D-meson
decays to search for the violation of the $|\Delta I|=1/2$ rule
if there is a sizable number events for other reactions including
$D^+\to \bar{K}^0 +\pi^0 + l + \nu_l$, but this need be pursued after
a compelling analysis of presently available data for the
determination of phase shifts of interest.
For a recent discussion on $|\Delta I|=3/2$ amplitudes,
see ref.~\cite{hep-ph/0506117}.

\bigskip

\noindent{\bf 7.} In summary, we point out that the FOCUS
collaboration with its large sample of $D_{l4}$ decays can
carry out a determination of much sought after $\pi K$ phase
shifts by adopting the methods of Pais and Treiman, and those of
Cabibbo and Maksymowicz, and Berends, Donnachie and Oades, and
go beyond establishing an interference phenomenon.  This would
be a valuable source of information for important low energy
observables such as pion-Kaon scattering lengths.

\bigskip
\noindent {\bf Acknowledgements:}  
It is a pleasure to thank Prof. G. Colangelo for suggesting
this investigation and for invaluable discussions.
BA thanks Prof. G. Dosch for a conversation, 
Prof. G. Kramer for a useful 
clarification, and the hospitality of the Theory Group,
Thomas Jefferson National Accelerator Facility,
USA at the time this work was initiated.  His work is supported in 
part by the CSIR under scheme number 03(0994)/04/EMR-II. 
It is a pleasure to thank Prof. D. Kim for comments on
the first version of the paper.

\end{document}